\documentclass[aps,twocolumn,prl,floatfix]{revtex4}
\usepackage[pdftex]{color}
\usepackage{latexsym,hyperref,amssymb,amsmath,gensymb}
\usepackage{graphicx,kec}
\usepackage{makeidx}

\newcommand{\om}{\omega}
\newcommand{\by}{\times}

\newcommand{\btab}{\begin{tabular}}
\newcommand{\etab}{\end{tabular}}
\newcommand{\barr}{\begin{array}}
\newcommand{\earr}{\end{array}}
\newcommand{\bpm}{\begin{pmatrix}}
\newcommand{\epm}{\end{pmatrix}}
\newcommand{\bit}{\begin{itemize}}
\newcommand{\eit}{\end{itemize}}
\newcommand{\ben}{\begin{enumerate}}
\newcommand{\een}{\end{enumerate}}
\newcommand{\bct}{\begin{center}}
\newcommand{\ect}{\end{center}}

\newcommand{\lt}{\left}
\newcommand{\rt}{\right}

\definecolor{navyblue}{rgb}{.05,0,.55}

\begin{document}
\title{Fast Light, Fast Neutrinos?}
\date{\today}
\author{Kevin Cahill}
\affiliation{Department of Physics \& Astronomy,
University of New Mexico, Albuquerque, New Mexico\\
Physics Department, Fudan University, Shanghai, China}
\email{cahill@unm.edu}
\begin{abstract}
In certain media, light has been observed with group velocities faster than the speed of light.   The recent OPERA report of superluminal 17 GeV neutrinos may describe a similar phenomenon.
\end{abstract}
\maketitle
Over the past decade,
Boyd and others have observed 
light moving through certain media
with group velocities
faster than the speed of light in 
vacuum~\cite{Boyd2003,Neifeld2003,Brunner2004}\@. 
The OPERA collaboration~\cite{Opera2011}
may have detected the neutrino
analog of this ``fast light'' phenomenon.
\par
The group velocity of an amplitude
\beq
A(x,t) = 
\int e^{i(k \cdot x - \omega t)} 
\, B(k) \, dk
\label {amplitude}
\eeq
is determined by the condition that
the phase remain constant
as a function of wave-number,
\( v_g = d \om/dk \)\@.
When neutrinos traverse a medium with
a complex index of refraction 
\(n\), scattering makes
the wave number \(k\) 
complex with a positive imaginary part.
The real frequency \(\omega\) 
is related to the complex
wave number \(k\) by
\(k = n \, \om / c\)\@.
If \(n_r\) is the real part
of the index of refraction,
then the group velocity is
\beq
v_g = \frac{d\omega}{dk_r}
= \frac{c}{n_r + \omega \, 
dn_r/d\omega}.
\label {PGV9}
\eeq
\par
The index of refraction \(n\)
is related to the forward scattering
amplitude \(f\) and the
density \(N\) of scatterers 
by~\cite{Liu1992}
\beq
n(\omega) = 1 + \frac{2\pi c^2}{\omega^2} 
N f(\omega)
\label {PGV12 again}
\eeq
in which for simplicity I replaced
\(k^2\) by \(\om^2/c^2\), which introduces
an error of less than \(10^{-19}\)
for 17 GeV neutrinos
with a mass of less than 
2 eV\(/c^2\)~\cite{Opera2011}\@.
\par
Group velocities faster than \(c\)
can occur when the frequency \(\omega\) 
is near a resonance in the total 
cross-section.
For instance, if the amplitude 
for forward scattering is
of the Breit-Wigner form
\beq
f(\om) = f_0 \,\frac{\Gamma/2}
{\om_0 - \om - i \Gamma/2}
\label {BR}
\eeq
then the real part of the 
index of refraction is
\beq
n_r(\omega) = 1 + 
\frac{\pi c^2 N f_0 \Gamma (\om_0 - \om)}
{\om^2 \lt[(\omega - \omega_0)^2 
+ \Gamma^2/4\rt]}
\label {n_r}
\eeq
and by (\ref{PGV9}) 
the group velocity is
\beq
v_g  = c \, \lt[ 1 + 
\frac{\pi c^2 N f_0 \Gamma \, \om_0}
{\om^2}
\, \frac{\lt[ 
\lt( \omega - \omega_0 \rt)^2 - \Gamma^2/4 \rt]}
{\lt[ (\omega - \omega_0)^2 + \Gamma^2/4\rt]^2} 
\rt]^{-1}  
\label {real slow/fast light}
\eeq
which is superluminal 
if \(( \om - \om_0)^2 < \Gamma^2/4\)\@.
\par
More generally, we may use 
the optical theorem and the
regularized Kramers-Kronig formula
\beq
n_r(\omega) = 1 + 
\frac{cN}{\pi}
\int_0^\infty \frac{
\sigma_{\mbox{\scriptsize{t}}}(\omega') 
- \sigma_{\mbox{\scriptsize{t}}}(\omega)}
{\omega^{\prime 2} - \omega^2}
\, d\omega'
\label {n_r by Kr-Kr regularized}
\eeq
in which \(\sigma_{\mbox{\scriptsize{t}}}\)
is the total cross-section
to write the group velocity 
in terms of the principal part
of an integral
\beq
\frac{c}{v_g(\om)} =  
1 + \frac{cN}{\pi} P \!\!
\int_0^\infty \!
\frac{\lt[\sigma_{\mbox{\scriptsize{t}}}(\omega') 
- \sigma_{\mbox{\scriptsize{t}}}(\omega)\rt](\om^{\prime 2} + \om^2)}
{(\omega^{\prime 2} - \omega^2)^2}
\, d\omega' 
\label {v_g}
\eeq
which shows the effect of
scattering on group velocities.
Just as the scattering of photons
by atoms can cause 
fast~\cite{Boyd2003,Brunner2004}, 
slow~\cite{Hau1999},
and even backward~\cite{Boyd2006}, 
light, 
so too the scattering of neutrinos by electrons
and quarks may make neutrino
group velocities that are faster
or slower than the speed of light.
The \(\nu_\mu\)-nucleon
charged-current total cross-section 
rises linearly up to 
300 GeV~\cite{Nakamura2010} and makes
a positive contribution to the 
integral (\ref{v_g})\@.
Yet the OPERA Collaboration~\cite{Opera2011}
may have discovered 
``fast neutrinos''---neutrinos 
with group velocities
faster than the speed of light~\cite{Vladan2011}\@.
Their high group velocity 
\((v-c)/c = 2.48 \by 10^{-5}\)
may arise from a resonance 
in neutrino-electron
and/or neutrino-quark scattering at
an energy \(\om_0\) somewhere
near 17 GeV\@.
%See however~\cite{Glashow2011}\@.
\par
A group velocity faster than \(c\)
doesn't violate special relativity,
but a superluminal signal velocity
would~\cite{Neifeld2003,Brunner2004}\@.
\begin{acknowledgments}
I should like to thank Franco Giuliani, Lei Ma,  
and David Waxman
for helpful conversations.
\end{acknowledgments}
\bibliography{physics}
\end{document}